\def\epsfpreprint{Y}   

\if \epsfpreprint Y \input epsf \fi

\def\myfigure#1#2#3{\if \epsfpreprint Y
\begin{figure}
\epsfxsize=#3truein
\epsffile{fig#1.eps}
\vbox{\parindent=0pt\baselineskip=12pt
Figure #1. #2}\label{F#1}\end{figure} \fi}
\def\captionone{$m_H/f_\pi$ vs. $m_H$ for the na\"\i ve ($\beta_2=0$)
lattice action. The solid line is the large $N$ result scaled to $N=4$.
The boxes are results from reference 1, the diamonds and crosses
are numerical results from references 4.}
\def\captiontwo{Leading order cutoff effects in the  invariant
$\pi-\pi$ scattering amplitude at $90^0$ for the na\"\i ve action
($\beta_2=0$) and for the action with
a four derivative term turned on to maximal allowed strength
($\beta_2=-\beta_{2,t.c.}$). The
dotted line represents center of mass energies $W=2M_H$, the dashed line
$W=3M_H$ and the solid line $W=4M_H$.
}
\def\captionthree{Preliminary Monte Carlo data on $F_4$ with a more complicated
action which yields a higher Higgs mass.}
\def\captionfour{Generic phase diagram containing qualitative features common
to all models investigated.}

\def\spose#1{\hbox to 0pt{#1\hss}}
\def\ltapprox{\mathrel{\spose{\lower 3pt\hbox{$\mathchar"218$}}
 \raise 2.0pt\hbox{$\mathchar"13C$}}}
\def\gtapprox{\mathrel{\spose{\lower 3pt\hbox{$\mathchar"218$}}
 \raise 2.0pt\hbox{$\mathchar"13E$}}}
\def\inapprox{\mathrel{\spose{\lower 3pt\hbox{$\mathchar"218$}}
 \raise 2.0pt\hbox{$\mathchar"232$}}}

\def\mass{{M_{\sigma}}}
\def\batio{{{M_{\sigma}}\over{\Lambda}}}
\def\fhi{\vec \phi}
\def\ratio{{{M_{\sigma}}\over{F}}}
\documentstyle[ichep,12pt]{article}
\title{\normalsize THE TRIVIALITY BOUND ON THE HIGGS MASS; ITS VALUE AND WHAT
IT MEANS.
\thanks{ Research partially supported under DOE grants \# DE-FG05-90ER40559
and \# DE-FG05-85ER250000.}}
\author{Herbert Neuberger\\Department or Physics and Astronomy,
Rutgers University\\Piscataway, NJ 08855-0849, USA.\\ \\
Urs M. Heller, Markus Klomfass, Pavlos Vranas\\Supercomputer
Computations Research Institute, The Florida State University\\Tallahassee, FL
32306, USA.
}

\begin{document}
\finalcopy

\maketitle

\abstract{Older lattice work exploring the Higgs mass triviality bound
is briefly reviewed. It indicates that a strongly interacting scalar
sector in the minimal standard model cannot exist; on the other hand
low energy QCD phenomenology might be interpreted as an indication
that it could. We attack this puzzle using the $1/N$ expansion and
discover a simple criterion for selecting a lattice action that is
more likely to produce a heavy Higgs particle. Depending on the
precise form of the limitation put on the cutoff effects, our large
$N$ calculations, when combined with old numerical data, suggest that
the Higgs mass bound might be around 750 $GeV$, which is higher than
the $\sim 650~GeV$ previously obtained.  Preliminary numerical work
indicates that an increase of at least 19\% takes place at $N=4$ on
the $F_4$ lattice when the old simple action is replaced with a new
action (still containing only nearest neighbor interactions) if one
uses the lattice spacing as the physical cutoff for both actions.  It
appears that, while a QCD like theory could produce $M_H / F ~ \sim
6$, a meaningful ``minimal elementary Higgs'' theory cannot have
$M_H/ F~ \gtapprox 3$. Still, even at 750 $GeV$, the Higgs particle is so
wide ($\sim 290~$GeV), that one cannot argue any more that the scalar
sector is weakly coupled.} 

\vskip-1pc
\onehead{OVERVIEW.}
The aim of this section is to communicate the logical framework of
``triviality'' in as precise a fashion as possible and restrict the
presentation  of details to only the most important numbers. We believe that
the logical framework has reached maturity but the specific numbers might still
fluctuate by several percent before settling down within a year or so.

 Consider a model with a scalar field transforming under an internal $O(4)$
symmetry
\break\newpage\vglue2pt
\noindent group. The system is in the broken phase with three pions ($\pi$) and
one unstable massive particle, denoted by $\sigma$ or $H$. The pion decay
constant is denoted by $F$ ($F=246~GeV$) and the scalar selfcoupling is defined
by
 $$
 g={{3M_{\sigma}^2}\over{F^2}},\eqno{(1)}$$
 where $M_\sigma$ is the real part of the complex pole representing the Higgs
particle. The model also has an ultraviolet cutoff $\Lambda$.

 When $g$ is small we have
 $$
 {{M_{\sigma}}\over {\Lambda}}\approx C g^{-{{b_2}\over {b_1^2 }}}
e^{-{1\over{b_1 g}}} (1+ O(g)),\eqno{(2)}$$
 where
 $$
 b_1 = {4\over{(4\pi )^2}}~~~~~~~~~~b_2=-{{26}\over{3(4\pi )^4}}\eqno{(3)}$$
 $C$ is a finite but unknown constant.

 If we try to take $\Lambda\rightarrow\infty$ at fixed $\mass$ and equation (2)
applies, $g$ must go to zero and hence ${{\mass}\over F}
\rightarrow 0$, implying a vanishing physical mass for $H$. Conversely, if we
increase $g$ and ignore the eventual large corrections to (2), then $\batio$
increases. We cannot tolerate increases of $\batio$ beyond some number of order
1 (${1\over 2}$ say) because cutoff effects will become all important at
energies $E\sim 2\mass$ (say) and our model will loose
the predictive power typical
of renormalizable theories.

 Two obvious improvements need to be made to turn the above into a quantitative
estimate for $g_{\rm max}$ and, by definition, for the ``triviality'' bound on
the Higgs mass.

 $\bullet$ For small $g$ we need the number $C$.

 $\bullet$ For larger $g$ we need to replace (2) by a more accurate relation.

 Both improvements require nonperturbative calculations and were implemented in
1988-90 by lattice field theory methods in some
 particularly simple models using

 1. strong coupling and renormalization group improved perturbation theory on
hypercubic lattices$^1$.

 2. Monte Carlo on hypercubic lattices$^2$.

 3. Monte Carlo on the $F_4$ lattice$^3$.

 In all models the simplest discretization of a $|\phi |^4$ theory was adopted
and the bounds were obtained in the nonlinear field limit. In the hypercubic
work the result was $M_H \le 650~GeV$ if ${{M_H}\over {\Lambda}}\equiv M_H a <
{1\over 2}$ ($a$ is the lattice spacing), a result recently updated$^4$
upwards by about 5-7\% . The $F_4$ lattice was investigated because it differs
from the hypercubic lattice in that it does not suffer
from contamination by Lorentz
symmetry violations at order ${1\over {\Lambda^2}}$ and yielded $M_H \le
600~GeV$ at cutoffs similar to the ones where the bound was obtained on
hypercubic lattices. On the basis of these numbers, which amount to
a $g$ of about ${2\over 3}$ of the unitarity bound$^1$,
a general feeling
ensued that strong scalar selfcouplings in the minimal standard model are
excluded by triviality.

 So far we discussed the so called ``obvious'' improvements. One needs to take
the analysis further by two additional steps:

 $\bullet$ The vague requirement that ${{M_H}\over {\Lambda}}$ not become too
large must be turned into a precise, regularization independent restriction.
First steps in this direction were taken in references 1,3 and 4.

 $\bullet$ One needs to gain an understanding of how the triviality bound
obtained under a given physical restriction depends on the precise details of
the cutoff scheme. There was an awareness to the issue already in 1987$^5$
and
some exploratory study was carried out in 1988$^6$; also the
work in reference
3 had some relevance to this issue. However, a serious investigation was
started only in 1991$^7$ and is still continuing.
More recently, additional
results relevant to this were presented in reference 4.

 Again, both of the above steps are of a nonperturbative nature. Although the
calculation of cutoff effects and the imposition of the related restriction can
be carried out in the loop expansion, only tree level results in particularly
simple cases are known and the reliability of the expansion is not at all
clear. The second step is essential because without it we cannot infer a bound
on the mass of the ``real'' Higgs particle; assuming that the minimal standard
model applies in some energy range we know that there has to be a cutoff
$\Lambda$ but it is provided by an embedding theory of an unknown nature.

 We are only interested in the situation where ${{M_H}\over{\Lambda}}$ is
sufficiently small so that any process at energy $E \ltapprox
4 M_H$ has only small
${1\over{\Lambda^2}}$ corrections. We admit some ``fine tuning'' in the sense
that ${{M_H}\over
 {\Lambda}}$ is small, but since we are looking for an upper bound on
$M_H$, ${{M_H}\over {\Lambda}}$ will never be extremely small. However, we
do exclude any excessive ``fine tuning'' that would make some of the generic
${1\over{\Lambda^2 }}$ corrections disappear. In other words, the scalar sector
is assumed to be representable to order ${1\over{\Lambda^2}}$ by an effective
action$^8$
 $$
 L_{\rm eff} =L_{\rm ren} +{1\over{\Lambda^2}} \sum_A c_A {\cal O}_{A}
,\eqno{(4)}$$
 where the sum over $A$ is finite, ${\rm dim} {\cal O}_A \le 6$ and order
${1\over{\Lambda^4}}$ terms are considered negligible. The set of operators $\{
{\cal O}_A \}$ is restricted by symmetry requirements and the coefficients
$c_A$ depend on the details of the embedding theory.

 The cutoff $\Lambda$ is taken as some well defined quantity in the particular
regularization, for example $\Lambda =a^{-1}$ where $a$ is the length of a bond
on the hypercubic lattice. The operators ${\cal O}_A$ are normalized in some
reasonable way, for example, demanding the matrix elements of ${\cal O}_A$ to
be of order 1 between states of energy $E\sim M_H$.

 The restriction on the cutoff effects translates into some limits on the
$c_A$'s. If we vary the bare parameters in a given regularization, subjected to
the above constraints on the $c_A$'s, a maximal value for $M_H$ ensues,
typically at points where the limits on the $c_A$'s are saturated. Note that
this process of extremization is highly nonlinear in the bare parameters and
that this nonlinearity cannot be eliminated by RG improvement.

 We do not know what the true embedding theory
is\protect\footnote{We also don't know that one
at all exists for the minimal standard model with an elementary Higgs, but this
we have assumed.} but whatever its nature might be it will
generate some $M_H$ and
some $c_A$'s . The same $M_H$ and the same $c_A$'s can be generated by almost
any reasonable cutoff model; such a model would agree with the true theory in
the sense that it will generate the same physical effects as equation (4) does
to order ${1\over{\Lambda^2}}$.

 Therefore, if we look at a large enough class of cutoff models it will
quite likely
contain a representative of the true embedding theory at the level of equation
(4). The bound on $M_H$ in this ``sufficiently large'' class of theories will
also have to be obeyed  by the ``true'' Higgs particle.

 What is a ``sufficiently large'' class of theories? The minimal requirement is
that it contain a sufficient number of free parameters to vary all the $c_A$'s
independently. The numerical data available to date does not yet satisfy this
minimal requirement.

 A further refinement enters at this point: We do not care about all the
independent $c_A$'s that can appear in equation (4) because some combinations
of $c_A$'s do not enter the S-matrix to order ${1\over{\Lambda^2}}$.
Eliminating the redundant operators, and absorbing the $c_A$'s that correspond
to operators of dimension $\le 4$ we are left with two measurable
$c_A$'s that parametrize all observable cutoff effects at order
${1\over{\Lambda^2}}$.

 Just counting parameters isn't yet quiet satisfactory; the bound depends on
the
bare cutoff action in a highly nonlinear way and we may worry that the range of
allowable magnitudes for the $c_A$'s is limited by some restriction on the bare
action, like, for example, that the Hamiltonian is bounded from below. One
needs therefore to carry out some experimentation before a responsible estimate
for the triviality bound can be obtained. It also helps to develop a physical
intuition for which class of actions will be more likely to generate larger
Higgs masses.

 If we generalize the model to an $O(N)$ model and take $N$ to infinity it
turns out that restrictions on the cutoff effects in  the invariant amplitude
for $\pi$--$\pi$ scattering translate into a limitation on a single parameter
$c$ thus reducing effectively the number of measurable parameters $c_A$ by one.
The situation becomes as simple as it only could get and a comprehensive
analysis of the bound was carried out recently in this framework$^9$.

 Combining the results from that investigation with presently available $N=4$
numerical data, rough estimates for the ${1\over N}$ corrections were obtained
and approximate values for $N=4$ bound were derived. For the Higgs mass it was
found that a not too conservative bound is $M_H \le 750~GeV$ with an
expected accuracy of about 7\%\protect\footnote{Now $g\sim$90\% of the
unitarity bound.}. At 750 $GeV$ combining $N=\infty$
 results with a rough guess for the finite $N$ correction one gets an estimate
for the width of 290 $GeV$ significantly larger than the
 tree level result of 210 $GeV$.

 This indicates that triviality does not quite exclude strong scalar
selfinteractions in the minimal standard model. Still, triviality does rule out
anything even remotely close to a QCD like theory where $\ratio \sim 6$; for
example there is no known instance\protect\footnote{Preliminary
numerical data obtained
for lattice Yukawa models seem to indicate higher values of $\ratio$ but the
``unwanted'' particles there induce potentially large and yet unknown
cutoff effects.} of a regularized reasonable $|\phi |^4$ theory with $\ratio
=4$ and cutoff effects no larger than about 4\% on $\pi$--$\pi$ scattering at
CM energies of up to $2\mass$.

\onehead{SOME EXAMPLES.}
In this section several explicit examples of
both numerical and analytical results will be described in some
detail.

The typical result of a Monte Carlo or strong coupling analysis will look like
the discrete points in Figure 1.
For comparison we also show there the $N=\infty$ result rescaled to $N=4$. The
horizontal axis, labeled by $m_H$ gives the Higgs mass in lattice units,
i.e. $m_H ={{M_H}\over{\Lambda}}=aM_H$. Typically, cutoff effects on pion
scattering become of the order of a few percent when $m_H \approx .5$. The
vertical axis then gives the physical Higgs mass in units of $F=246~GeV$.
The parameter $\beta_2$ mentioned in the figure is a new parameter which when
set to zero implies an action of the simplest type; such actions have been
already quite thoroughly investigated.

\myfigure{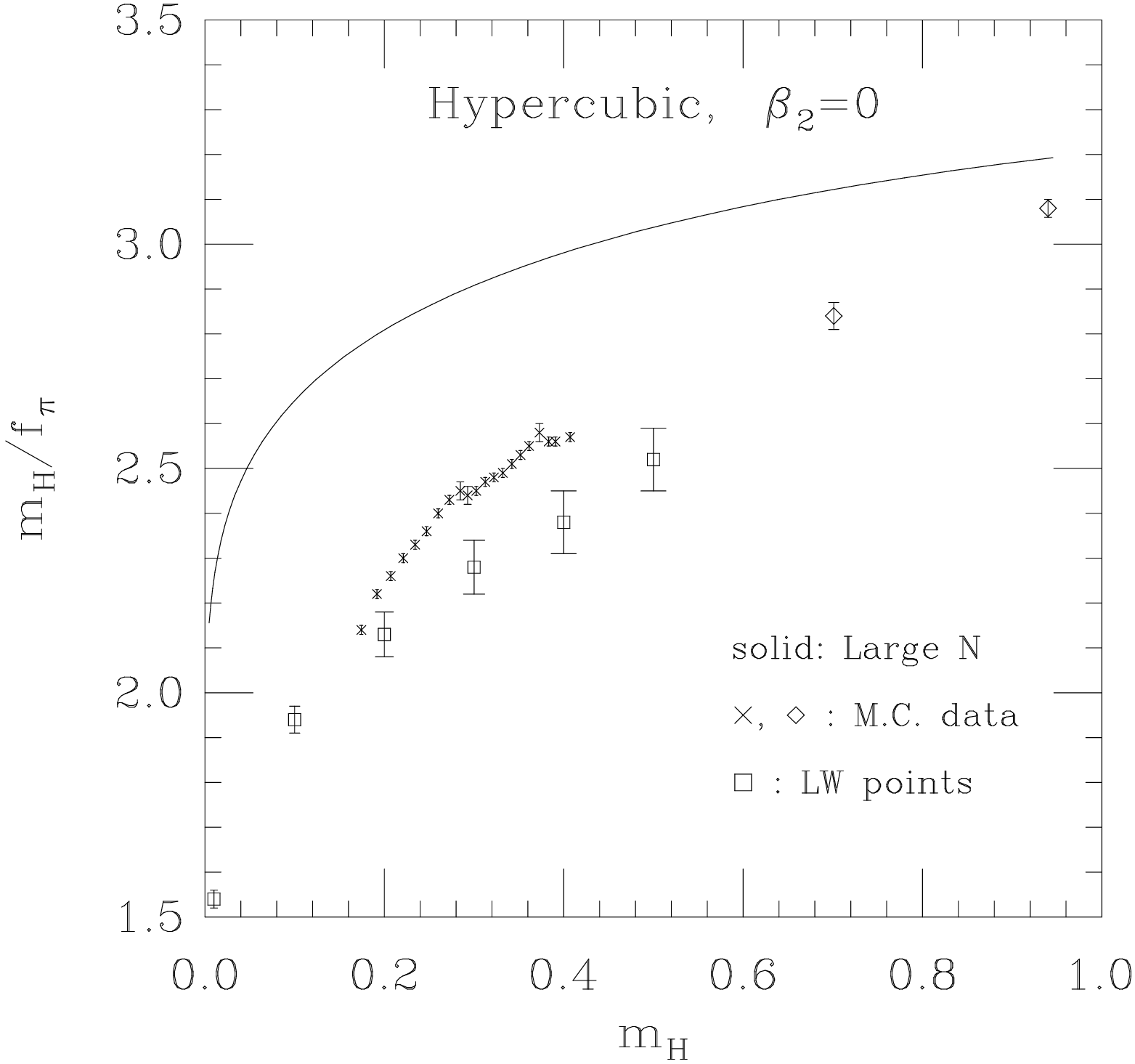}{\captionone}{3.8}

The measurements entering the creation of graphs like in Figure 1. deal with
two quantities: $f_{\pi} =F/\Lambda$ and $m_H =M_H /\Lambda$. The pion decay
constant in lattice units is, in turn, determined by two other quantities:
the wave function normalization of the pions, evaluated at zero
momentum, and the vacuum expectation value of the order parameter.

The
wave function normalization constant for the pions is typically determined
by fitting the lattice pion propagator. The vacuum expectation value
is extracted by direct measurement of the expectation value of the square
of the order parameter and extrapolated  to infinite volumes
using soft pion induced finite volume corrections. The method also provides a
check on the
wave function normalization constant.

The mass is extracted by direct
measurement of correlation decay in the imaginary time direction; this
works in spite of the instability of the scalar due to kinematical
constraints holding in sufficiently small volumes: The small volume
makes the lightest pion pair heavy enough to prohibit the decay. This
method is obviously imperfect and one needs independent checks on its
results.

Typical checks are done by fits to the scalar propagator or by use
of results
of $\beta$--functions to extrapolate across the transition from the
symmetric phase where measurements are cleaner conceptually and the
strong coupling expansion is also available. The first nonperturbative results
were obtained with this latter method${}^1$. Note that
the finite volume also obscures the difference between the pion components and
the scalar components and the field representatives of these particles have to
be chosen with some care. The method of reference 1 never needs to employ
finite volumes and agreement with its results provides a check on
the validity of the infinite volume extrapolation one carries out in the
Monte Carlo data analysis.

There is also a possibility to replace the measurement of $m_H$ by the
measurement of a differently defined selfcoupling which is more
convenient${}^8$. This leaves one with the problem to connect that coupling to
$m_H / f_\pi$ which might be attempted in perturbation theory${}^4$.
To date, the single method that seems to be able to allow (indirect) mass
measurements in the case where the Higgs particle is expected to be really wide
is that of reference 4.
With all methods it is the overall consistency of the numbers one obtains that
adds up to a believable result rather than the strength of one particular
approach over others.

Cutoff effects are calculated analytically; in Figure 2. we show an example
computed in the $1/N$ expansion for the $F_4$ lattice.
Tree level estimates, where known, behave
similarly. The plotted
quantity is the percentage cutoff effect on the invariant $\pi$--$\pi$
scattering amplitude squared at various ratios of the center of mass energy to
the Higgs mass.

There are particular simplifications that occur at infinite $N$ that make the
calculation easier. One of these simplifications is that at infinite $N$ one
can write down nontrivial universal expressions (for $\pi$--$\pi$ scattering
for example) that have no cutoff dependence at all but are nonperturbatively
dependent on the coupling constant. Triviality tells us that these expression
cannot be made to represent the finite cutoff theory more and more accurately
with an increasing cutoff and this is seen in the explicit formulae relating
the bare parameters to the renormalized ones.

\myfigure{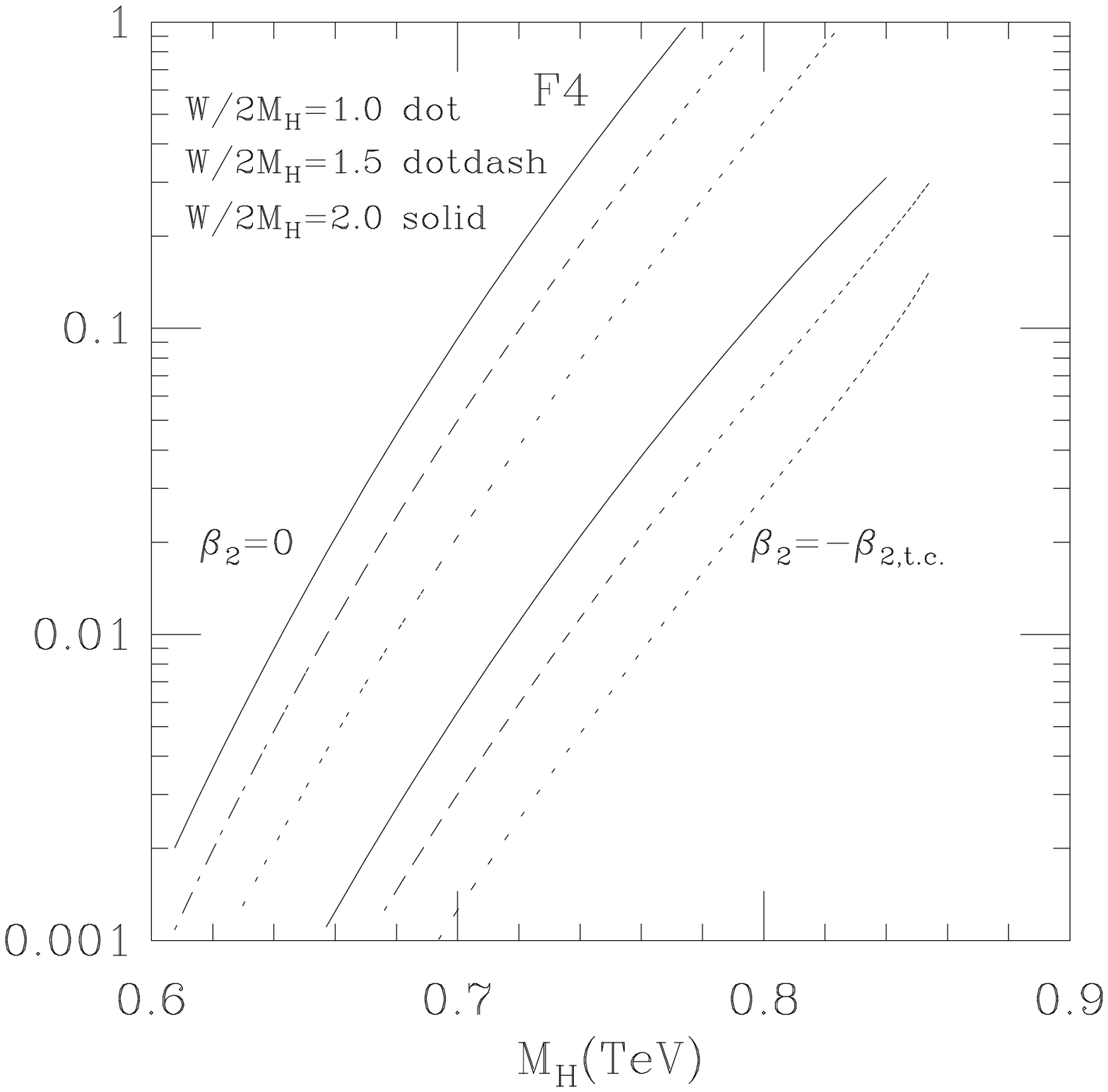}{\captiontwo}{3.8}

The so called universal expressions also know about triviality, and display
unphysical tachyonic poles at energies of the order of the cutoff, again
signaling the poorness of the approximation at higher energies. At low energies
the universal expressions can be used however to unambiguously separate out the
cutoff contribution to observables. Explicit calculation of these
cutoff effects to order $1/\Lambda^2$ to $\pi$--$\pi$ scattering yields
formulae that neatly factorize for all momenta into one parameter that depends
on the bare action, but not on the coupling, and a universal factor that
carries all the dependence on the momenta and couplings. This factor is common
to all regularizations and we have checked several. The factorization is clear
if one measures all momenta in units of one scale, the pion decay constant
being the most natural choice.

{}From Figure 2. we see that the cutoff effect increases when the model is
required to hold in a larger energy range. The figure also shows that by
changing the bare action the cutoff effects at fixed physical Higgs mass can be
decreased, increasing the triviality bound. The cluster of lines labeled by
$\beta_2 =0$ corresponds to
the simplest action for which numerical information is already available.

The horizontal
line is in units of the physical Higgs mass and we see that the $N=\infty$
number
at 4\% violation is $\sim 680~GeV$ with the simplest action. However, when we
compare a graph like the one in Figure 1. to the data, we see that, at fixed
mass in lattice units ($m_H$) the large $N$ result has a tendency to
overestimate the $N=4$ numbers; this has to be taken into account and is
supposed to reduce the bound for the simplest action from $680~GeV$ at
$N=\infty$ to $600~GeV$ at $N=4$.

The
most important thing to pay attention to is the amount of increase induced
when the action is varied. One sees from Figure 2. that an increase of about
$80~GeV$ in the bound is expected at $N=\infty$ and this is
approximately confirmed at $N=4$ by our preliminary numerical work. Our
preliminary data is presented in Figure 3..

\myfigure{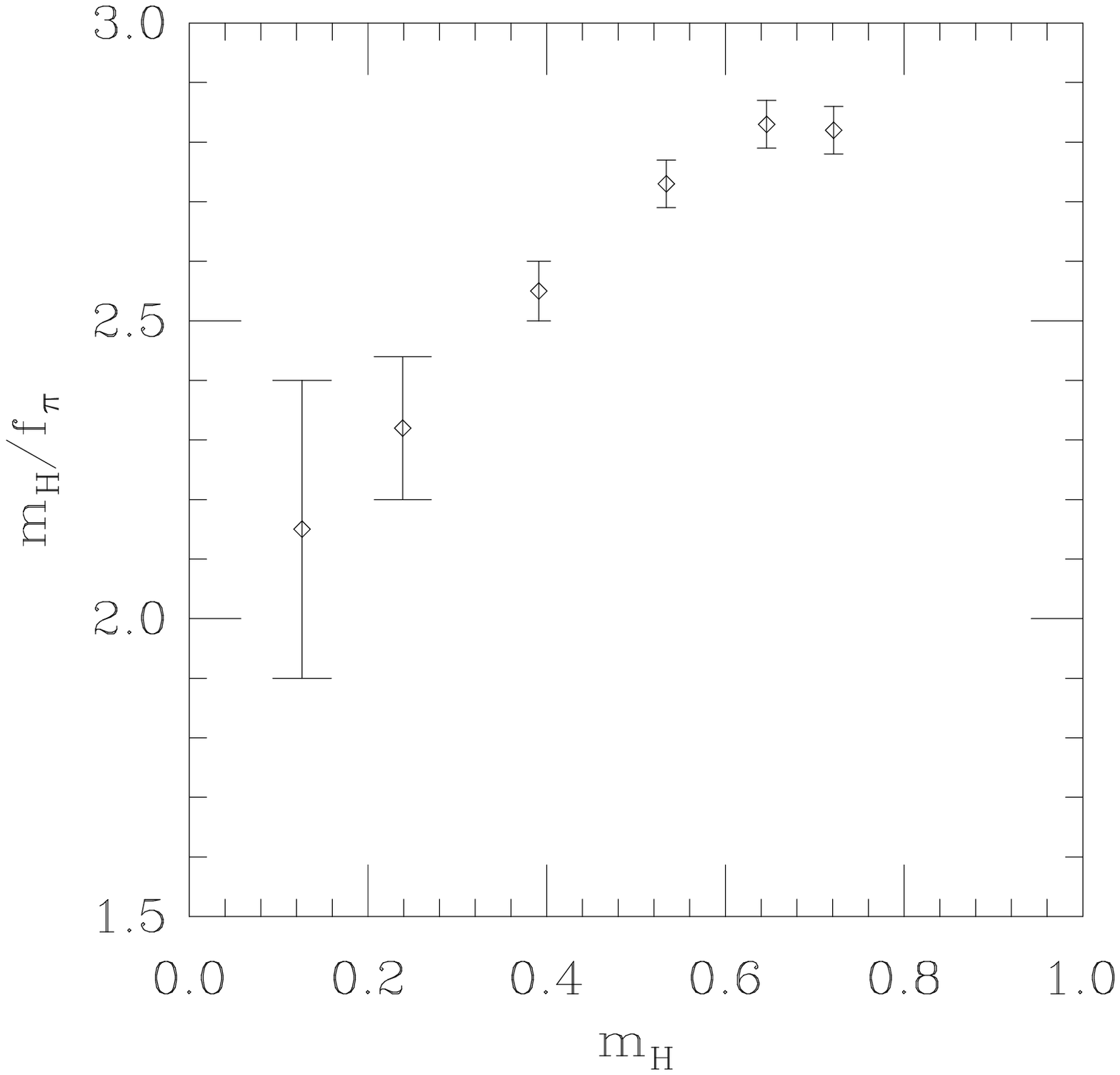}{\captionthree}{3.8}

Variations between different actions are predicted by large
$N$ to within less than a factor of 2 and always with the correct sign. This
can be checked in several cases already and is useful to make projections of
future results and guide the decision for what to simulate.

An important ingredient of the large $N$ study is that it helped us develop an
intuitive understanding for the way the bound depends on the bare action.
Firstly the bound appears always to be attainable in the nonlinear limit;
this was already observed with the simplest actions and is not surprising
because the bare coupling is maximized in this case. When the regularized model
is nonlinear one can think
about the Higgs resonance as a loose bound state of two pions in an
$I=0$, $J=0$ state. Pions in such a state attract because superposing
the field configurations corresponding to individual pions makes the
state look more like the vacuum and hence lowers the energy.

To see what
the action tries to do expand it in slowly varying fields to the form
$$
S_c=\int_x ~ \bigg [ {1\over 2} \vec \phi (-\partial^2
+2 b_0 \partial^4 )\vec \phi - {b_1 \over {2N}} (\partial_\mu \vec \phi
\cdot \partial_\mu \vec \phi )^2
$$
$$
 - {b_2 \over {2N}} (\partial_\mu \vec
\phi \cdot \partial_\nu \vec \phi - {1\over 4} \delta_{\mu , \nu }
\partial_\sigma \vec \phi \cdot \partial_\sigma \vec \phi )^2
\bigg ],\eqno{(5)}$$
where $\fhi^2 = N\beta$.

There are four
control parameters in the effective action of equation (5) but one is redundant
since to
this order it can be absorbed into the others by a field
redefinition:  The parameter $b_0$ can be absorbed into $b_1$ and $b_2$
by:
$$ \fhi \rightarrow {{\fhi - b_0 \partial^2 \fhi }\over{\sqrt{\fhi^2
+ b_0^2 (\partial^2 \fhi )^2 -2b_0 \fhi\partial^2 \fhi }}}\sqrt{N\beta}
\eqno{(6)}$$

The four
derivative term in the action can add or subtract to the pion--pion attraction
in the $I=J=0$ state.
The smallest cutoff effects are obtained when the
coupling $b_1$ of the four derivative term is set so that the term
induces the maximal possible repulsion between the pions, postponing
the appearance of the Higgs resonance to higher energies. This
corresponds to trying to make $b_1$ as negative as possible; there
is a limitation in trying to do so because a too large $b_1$
may induce a tendency for translational invariance to break spontaneously
by overly enhancing some nonzero momentum mode of the field. This limitation
was taken into account implicitly in all cases that we investigated at
$N=\infty$.

This physical picture essentially describes the most relevant
gross features of the
phase diagrams of all the regularized models that we have studied at infinite
$N$. The generic structure is displayed in Figure 4. and the main feature
is that there is a tricritical point on the ordinary symmetry breaking line.
To find the bound one has to be in the vicinity of a second order transition
point that is as far away
as possible from this tricritical point. The
tricritical point is explained by corresponding to so much attraction between
the two pions in the $I=J=0$ state that they make a massless bound state, which
condenses and couples to the energy momentum tensor, thus playing the role of a
dilaton.

\myfigure{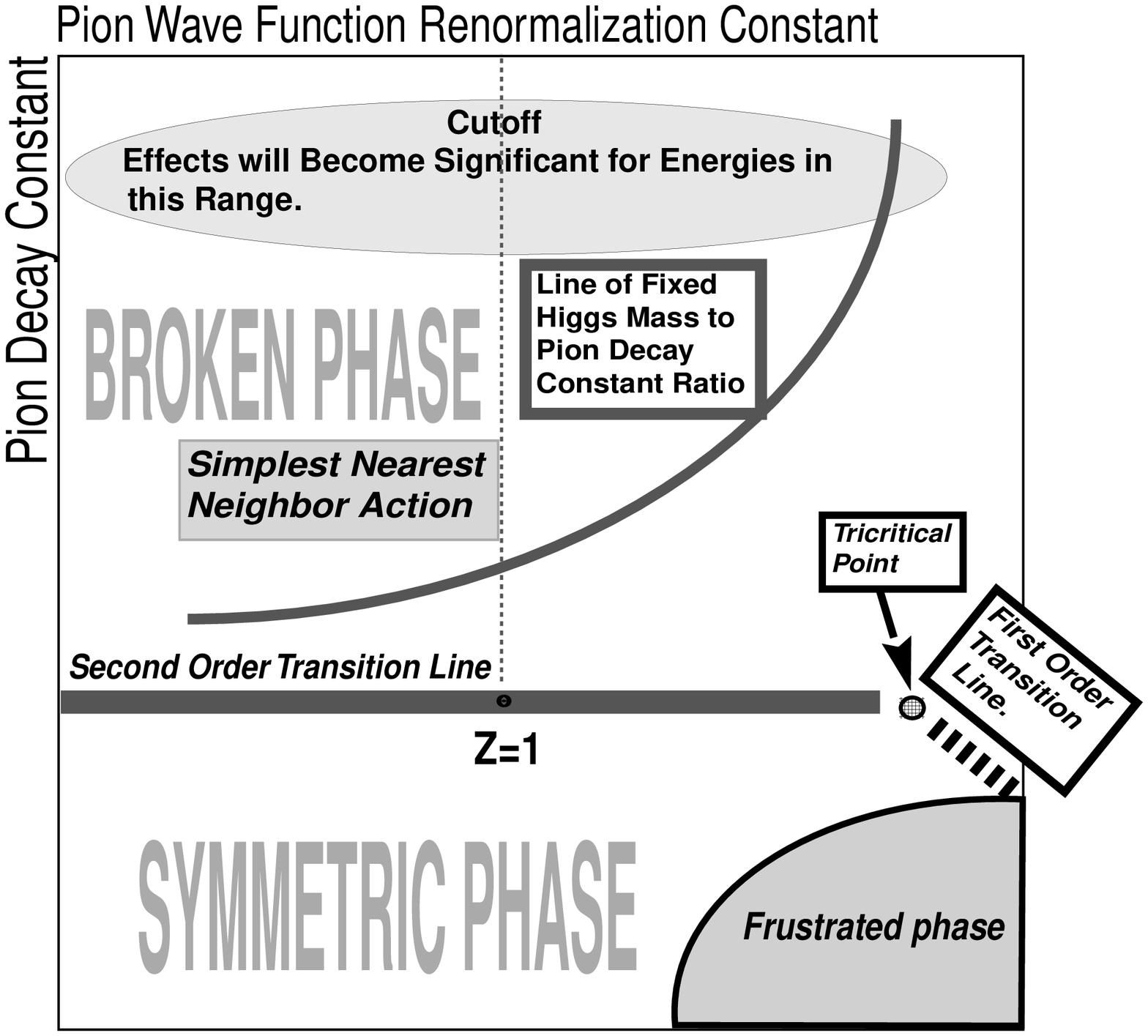}{\captionfour}{3.8}

 The tricritical point in the parameterization used in Figure 2. is
at $\beta_2 =\beta_{2,t.c.}$ just opposite in sign to the point where the
highest Higgs mass is obtained.

\if \epsfpreprint N

\onehead{FIGURE CAPTIONS}

\noindent {\bf Figure 1:} \captionone

\noindent {\bf Figure 2:} \captiontwo

\noindent {\bf Figure 3:} \captionthree

\noindent {\bf Figure 4:} \captionfour

\fi

\end{document}